\documentclass[11pt]{article}
\usepackage{cospar}
\usepackage{epsf}
\usepackage{url}
\usepackage{graphicx}

\usepackage[sectionbib]{natbib} %% Added 21-11-02
\title{NEUTRON STAR COOLING: THEORETICAL ASPECTS AND
       OBSERVATIONAL CONSTRAINTS}
\author{
D.~G.~Yakovlev\address{
Ioffe Physical Technical Institute,
Politekhnicheskaya 26, 194021 St.~Petersburg, Russia},
O.~Y.~Gnedin\address{ 
Space Telescope Science Institute,
3700 San Martin Drive, Baltimore, MD 21218, USA},
A.~D.~Kaminker$^1$, 
K.~P.~Levenfish$^1$, and 
A.~Y.~Potekhin$^1$
}
\begin{document}
\def\la{\;\raise0.3ex\hbox{$<$\kern-0.75em\raise-1.1ex\hbox{$\sim$}}\;}
\def\ga{\;\raise0.3ex\hbox{$>$\kern-0.75em\raise-1.1ex\hbox{$\sim$}}\;}

\maketitle

% **********************************************************************
\begin{abstract}
The cooling theory of isolated neutron stars is reviewed.
The main cooling regulators are discussed, first of all,
operation of direct
Urca process (or similar processes
in exotic phases of dense matter) and superfluidity
in stellar interiors. 
The prospects
to constrain gross parameters of supranuclear matter
in neutron-star interiors by confronting cooling
theory with observations
of isolated neutron stars are outlined.
A related problem of thermal states of transiently
accreting neutron stars with deep crustal heating
of accreted matter is discussed in application
to soft X-ray transients. 
\end{abstract}

% **********************************************************************
%                               TEXT BODY
% **********************************************************************
%%%%%%%%%%%%%%%%%%%%%%%%%%%% Sect. 1 %%%%%%%%%%%%%%%%%%%%%%%%%%%%%%%%%%%
\section*{INTRODUCTION}
Our knowledge of neutron star (NS)
interiors is still uncertain. For instance,
the fundamental problem of the equation of
state (EOS) at supranuclear densities
in NS cores is not solved.
Microscopic calculations are model dependent
and give a large scatter of possible EOSs
(e.g., Lattimer and Prakash, 2001), from stiff to soft
ones, with different compositions of
inner NS cores (nucleons, pion or kaon
condensates, hyperons, quarks). The
NS cooling theory enables one to study the internal structure
of NSs by confronting the cooling models
with observations.

NSs are born hot in supernova explosions,
with internal temperature $T \sim 10^{11}$ K,
but gradually cool down.
In about 30 s after the birth a star becomes 
transparent for neutrinos generated
in its interiors. In the
following neutrino-transparent stage
the cooling is realized via
neutrino emission from the entire stellar body
and via heat transport to the surface resulting
in the thermal emission of photons.

The foundation of the strict NS cooling theory was laid
by Tsuruta and Cameron (1966). The recent development of the
theory has been reviewed,
e.g., by Pethick (1992), Page (1998a, b), and Yakovlev et al.\ (1999).
Here, we discuss the current state of
the theory as confronted with observations
of isolated NSs and outline also a related problem
of accreting NSs with application
to soft X-ray transients.

% *************************************************************************
\section*{COOLING CALCULATIONS AND MAIN COOLING REGULATORS}
% *************************************************************************

The hydrostatic structure of NSs
can be regarded as temperature-independent.
For a given EOS of dense matter, one can
construct a sequence of NS models with
different central densities $\rho_{\rm c}$,
masses $M$ and radii $R$ 
($M$ means gravitational mass, and $R$ 
circumferential radius). For many EOSs, low-mass NSs 
(with $\rho_{\rm c} \la 2 \, \rho_0$, where
$\rho_0 \approx 2.8 \times 10^{14}$ g cm$^{-3}$
is the density of saturated nuclear matter) contain
the nucleon cores (neutrons, n, with an admixture of protons, p, 
electrons, e, and possibly
muons). Massive NSs have the inner cores whose
composition is very model dependent
(e.g., Lattimer and Prakash, 2001).

NS cooling is governed by general relativistic
equations (Thorne, 1977) of heat diffusion inside
the star.
Their solution gives
the distribution of the temperature $T$ inside the star
versus time $t$, and the effective
surface temperature $T_{\rm s}(t)$. The main
ingredients of the cooling theory are the neutrino emissivity
$Q$, heat capacity $C$ and
thermal conductivity of NS matter.

It is conventional (Gudmundsson et al., 1983)
to divide the NS into the interior and
the outer he\-at-blanketing envelope
which extends to the density 
$\rho_{\rm b}\! \sim\! (10^{10}\!-10^{11})$ g cm$^{-3}$
($\sim\! 100$ meters under the surface). The
thermal structure of the envelope
is studied separately in the stationary,
plane-parallel approximation to relate
$T_{\rm s}$ to the temperature
$T_{\rm b}$ at $\rho=\rho_{\rm b}$. The diffusion equations
are solved then at $\rho \geq \rho_{\rm b}$.

The NS thermal luminosity is
$ L_\gamma = 4 \pi \sigma R^2 T^4_{\rm s}(t)$.
Both, $L_\gamma$ and $T_{\rm s}$, refer to a locally-flat
reference frame on the NS surface. A distant
observer would register the ``apparent'' luminosity
$L_\gamma^\infty = L_\gamma (1 - r_{\rm g}/R)$
and the ``apparent'' effective temperature
$T_{\rm s}^\infty = T_{\rm s} \, \sqrt{1 - r_{\rm g}/R}$,
where $r_{\rm g}=2GM/c^2$ is the Schwarzschild radius.
If the surface temperature is nonuniform
(e.g., in the presence of a strong magnetic field),
$L_\gamma^\infty$ is determined by a properly
averaged surface temperature (e.g., Potekhin and Yakovlev, 2001).

The theory provides
{\it cooling curves}, $T_s^\infty(t)$, 
to be compared with observations.
One can distinguish three cooling stages:
the internal relaxation stage ($t \la 10-50$ yrs;
Lattimer et al., 1994; Gnedin et al., 2001),
the neutrino cooling stage (the neutrino luminosity
$L_\nu \gg L_\gamma$, $t \la 10^5$ yrs), and
the photon stage ($L_\nu \ll L_\gamma$,
$t \ga 10^5$ yrs). 
After the thermal relaxation, the redshifted
temperature
$ \widetilde{T}(t)= T(r,t)\; {\rm e}^{\Phi(r)}$
becomes constant throughout the stellar interior
($\Phi(r)$ being the metric function which determines
gravitational redshift), and the problem reduces to solving
the global thermal-balance equation
\begin{equation}
     C_{\rm tot}(\widetilde{T}) \, 
     {{\rm d} \widetilde{T} \over {\rm d} t}  =
      - L_\nu^\infty (\widetilde{T}) - L_\gamma^\infty (T_{\rm s}),
\qquad
     L^\infty_\nu (\widetilde{T}) = 
     \int   {\rm d}V \, Q (T) \, {\rm e}^{2 \Phi},
\qquad
     C_{\rm tot} (\widetilde{T}) = 
     \int {\rm d}V \, C(T),
\label{therm-balance}
\end{equation}
where ${\rm d}V$ is a proper volume element, 
$C_{\rm tot}$ is the total NS heat capacity, and
$L^\infty_\nu$ is the neutrino
luminosity as detected by a distant observer.

The main cooling regulators
at the neutrino cooling stage 
are: ({\it 1}) neutrino emission
in NS interiors,
and ({\it 2}) the effects of baryon superfluidity
on this emission.

%%%%%%%%%%%%%%%%%%%%%%%%%%%%%%%%%%%%%%
\newcommand{\rrr}{\rule{0cm}{0.3cm}}

\begin{table}[t]
\caption{Main processes of slow neutrino emission
in nucleon matter: Murca and
bremsstrahlung}
\begin{center}
  \begin{tabular}{|lll|}
  \hline
  Process   &    &  $Q_{\rm s}$, erg cm$^{-3 \rrr}$ s$^{-1}$ \\
  \hline
  Murca &
  ${\rm nN \to pN e \bar{\nu} \quad
   pN e \to nN \nu } $ &
  $\quad 10^{20 \rrr}-3 \times 10^{21}$  \\
  Brems. &
  ${\rm NN \to NN  \nu \bar{\nu}}$  &
  $\quad 10^{19 \rrr}-10^{20}$\\
   \hline
\end{tabular}
\label{tab-nucore-slow}
\end{center}
\end{table}

({\it 1}) The major neutrino mechanisms in
nucleon matter at $\rho \la 2 \, \rho_0$ are
modified Urca (Murca)
process and NN-bremsstrahlung (brems);
see Table 1, where N is a nucleon, n or p.
These mechanisms are relatively weak and produce
{\it slow} cooling. 
At higher $\rho$,
the neutrino emission can be strongly enhanced
by the onset of direct Urca (Durca) process 
(Lattimer et al., 1991) in
nucleon matter or similar (but somewhat weaker)
processes in exotic phases
of matter. These processes (Table 2) can
enhance the neutrino emissivity by 2--7 orders of magnitude and
lead to {\it fast} cooling.
An example, the enhancement by Durca process,
is shown on the right panel of Figure 1 and discussed
in the next section.
In nonsuperfluid matter,
the emissivities of slow and fast neutrino processes
can be written as
\begin{equation}
   Q_{\rm slow}= Q_{\rm s}\, T_9^8,\qquad
   Q_{\rm fast}= Q_{\rm f}\, T_9^6,
\label{Qnu}
\end{equation}
where $T_9=T/(10^9~{\rm K})$, while $Q_{\rm s}$ and $Q_{\rm f}$ are
slowly varying functions of $\rho$ (Tables 1 and 2).

({\it 2}) It is widely accepted that dense matter
can be superfluid (e.g., Lombardo and Schulze, 2001;
also see Yakovlev et al., 2001a, for references). Microscopic 
calculations predict superfluidity of neutrons in NS
crusts and cores, and superfluidity of
protons in NS cores; they predict also superfluidity
of hyperons or quarks. A superfluidity of any
baryon species is characterized by its own density
dependent critical temperature $T_{\rm c}(\rho)$
which is very model dependent and ranges from $\sim 10^8$ to
$\sim 10^{10}$ K. An exclusion is provided by pairing
of unlike quarks where
$T_{\rm c}$ can be as high as
$\sim 5 \times 10^{11}$ K. Any superfluidity
of baryons reduces neutrino
processes involving these baryons due to
an energy gap in the baryon dispersion relation.
When $T$ falls below $T_{\rm c}$, the superfluidity
initiates also a specific neutrino emission 
due to Cooper pairing of baryons (Flowers et al., 1976).
In addition, it affects baryon heat
capacity (e.g., Yakovlev et al., 1999), but 
at the neutrino cooling stage
this effect is less important 
than the effect on neutrino emission.

%%%%%%%%%%%%%%%%%%%%%%%%%%%%%%%%%%%%%%
\begin{table}[t]
\caption{Leading processes of fast
neutrino emission
in nucleon matter and three models of exotic matter}
\begin{center}
  \begin{tabular}{|lll|}
  \hline
  Model              & Process             &
        $Q_{\rm f}$, erg cm$^{-3 \rrr}$ s$^{-1}$ \\
  \hline
  Nucleon matter &
  ${\rm n \to p e \bar{\nu} \quad
   p e \to n \nu }$ & $\quad
  10^{26 \rrr}-10^{27}$  \\
  Pion condensate &
  ${\rm q \to q e \bar{\nu} \quad
   q e \to q {\nu} } $ & $ \quad
  10^{23 \rrr}-10^{26}$  \\
   Kaon condensate &
   ${\rm q \to q e \bar{\nu} \quad
    q e \to q \nu } $ & $ \quad
   10^{23 \rrr}-10^{24}$ \\
   Quark matter &
   ${\rm d \to u e \bar{\nu} \quad  u e \to d \nu } $ & $  \quad
   10^{23 \rrr}-10^{24}$ \\
   \hline
\end{tabular}
\label{tab-nucore-fast}
\end{center}
\end{table}

Furthermore, NS cooling can be affected 
by the thermal conductivity
in heat-blanketing envelopes which regulates
the relation between $T_{\rm s}$ and $T_{\rm b}$.
The conductivity is modified by the presence
of light (accreted) elements (Potekhin et al., 1997)
or by strong surface magnetic
fields (e.g., Potekhin and Yakovlev, 2001;
Yakovlev et al., 2002b, 
and references therein). These effects are 
weaker than those cited above.
The NS cooling may also be regulated by other factors,
particularly, by reheating mechanisms
which may operate in NS interiors, for instance,
due to frictional dissipation of differential rotation
or ohmic decay of magnetic fields.
Some of these mechanisms have been reviewed by Page
(1998a, b).
 
%%%%%%%%%%%%%%%%%%%%%%%%%%%%%%%%%%%%%%%%%%%% 
\section*{OBSERVATIONS}
%%%%%%%%%%%%%%%%%%%%%%%%%%%%%%%%%%%%%%%%%%%%

We will confront the cooling theory with 
observations of thermal emission from nine middle-aged isolated
NSs. Recently the observations
have been reviewed by Pavlov et al.\ (2002b).
The data are summarized in Table 3 and
displayed in Figures 2 and 3.
Two young objects (RX J0822--43, 1E 1207--52) are radio-quiet
NSs in supernova remnants; RX J1856--3754
is also a radio-quiet NS. The other objects, 
Crab, RX J0205+6449, Vela, PSR 0656+14,
Geminga, and PSR 1055--52, are observed as radio pulsars.
RX J0205+6449 and the Crab pulsar
are associated with historical supernovae and
their ages are certain.
For RX J0822--43,
we take the estimated age $t=2-5$ kyr of the associated
supernova (as can be deduced, e.g.,
from a discussion in Arendt et al., 1991) centered at
$t=3.7$ kyr (Winkler et al., 1988).
For 1E 1207--52, we adopt the range from the
standard age of the associated supernova ($\sim 7$ kyr)
to the four times longer age taking into account
slow spindown of NS rotation detected in X-rays
(Pavlov et al., 2002a). For Vela, we take the
age interval from the standard spindown age to
the age reported by Lyne et al.\ (1996).
The age of RX J1856--3754 has been revised recently
by Walter and Lattimer (2002) from the kinematical reasons;
the errorbar is chosen in such a way to
clearly distinguish the revised value from the previous one.
The ages of other NSs are the standard spindown
ages with an uncertainty by a factor of 2.

For two youngest sources, RX J0205+6449 and the Crab pulsar, 
no thermal emission has been
detected, but the upper limits on $T_{\rm s}^\infty$
have been established.
For the next three sources, the values of 
$T_{\rm s}^\infty$
are obtained from the observed X-ray spectra using
hydrogen atmosphere models. Such models are more consistent with other
information on these sources
(e.g., Pavlov et al., 2002b) than the blackbody model of
NS emission. On the contrary, for 
PSR 0656+14, Geminga and PSR 1055-52 we present
the values of $T_{\rm s}^\infty$ inferred using the blackbody spectrum
which is more consistent for these sources.
Finally, for RX J1856--3754 we adopt the values inferred
using the analytic fit with Si-ash atmosphere model of Pons et al.\
(2001).

\renewcommand{\arraystretch}{1.2}
\begin{table*}[!t]   % "*" ignores the twocolumn-format if adopted
\caption[]{Observational limits on surface temperatures of isolated 
neutron stars}
\label{tab-cool-data}
\begin{center}
\begin{tabular}{|| l | c | c | c | l ||}
\hline
\hline
Source & $t$ [kyr] & $T_{\rm s}^\infty$ [MK] &  Confid.\  & References   \\
\hline
\hline
PSR J0205+6449     & 0.82   & $<$1.1   &      & Slane et al.\ (2002)    \\
PSR B0531+21 (Crab)& 1      & $<$2.1   &      & Tennant et al.\ (2001)  \\
RX J0822--4300     & 2--5   & 1.6--1.9 & 90\% & Zavlin et al.\ (1999)   \\
%                                              Zavlin, Truemper, Pavlov 1999
1E 1207--52       & $\ga$7 & 1.1--1.5 & 90\% & Zavlin et al.\ (1998) \\
%                                       Zavlin, Pavlov, Truemper 1998
PSR 0833--45 (Vela)& 11--25  & 0.65--0.71 & 68\% & Pavlov et al.\ (2001)\\
%                                Pavlov, Zavlin,  Sanwal 2003 Bad Honnef
PSR B0656+14      &  $\sim$110 & 0.91$\pm$0.05  & 90\%  &
Possenti et al.\ (1996) \\
%              Possenti, Mereghetti, Colpi 1996
PSR~0633+1748 (Geminga) & $\sim$340 & 5.6 (+0.7,--0.9) & 90\% &
Halpern and Wang (1997) \\
RX~J1856--3754    & $\sim$500 & 0.52$\pm$0.07 & --  & Pons et al.\ (2002) \\
PSR~1055--52      & $\sim$530 & 0.82 (+0.06,--0.08)& 90\% &
Pavlov and Teter (2002)  \\
\hline
\end{tabular}
\end{center}
\end{table*}
\renewcommand{\arraystretch}{1.0}

%%%%%%%%%%%%%%%%%%%%%%%%%%%%%%%%%%%%%%%%%%%%%%%%%%%%%%%
\section*{COOLING OF NEUTRON STARS WITH NUCLEON CORES}
%%%%%%%%%%%%%%%%%%%%%%%%%%%%%%%%%%%%%%%%%%%%%%%%%%%%%%%

We start 
with the simplest composition of the NS cores
(n, p, and e).  Illustrative cooling curves are 
calculated with our fully relativistic
nonisothermal cooling code (Gnedin et al., 2001). 
In the NS cores, we use a stiff phenomenological EOS
proposed by Prakash et al.\ (1988) (their model I
with the compression modulus of saturated
nuclear matter $K=240$ MeV). The maximum-mass
configuration has $M=1.977\,{\rm M}_\odot$,
$R=10.754$ km, and $\rho_{\rm c}=2.575 \times 10^{15}$
g cm$^{-3}$. The Durca process is open at
$\rho \geq \rho_{\rm D}=7.851 \times 10^{14}$ g cm$^{-3}$.
The NS model with $\rho_{\rm c}=\rho_{\rm D}$
has $M=M_{\rm D}=1.358\,{\rm M}_\odot$ and $R=12.98$ km.

%%%%%%%%%%%%%%%%%%%%%%%%%%%%%%%%%%%%%%%%%%%%%%%%%%%%%%%%%%%%%%
\begin{figure}
\centering
\epsfxsize=18cm
\epsffile[45 185 575 415]{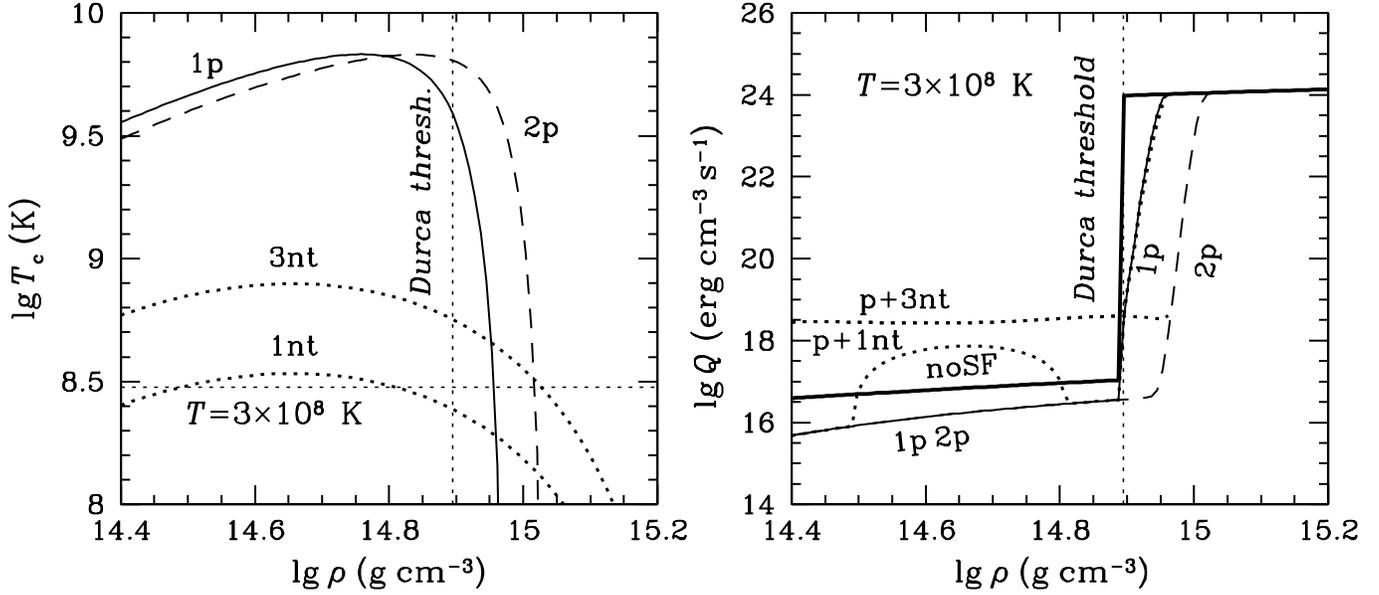}
\caption{
{\it Left panel}: density dependence of the critical temperatures of
superfluidity of protons (models 1p and 2p)
and neutrons (model 1nt and 3nt) in a NS core;
vertical dotted line is the Durca threshold;
horizontal line is the temperature 
$T=3 \times 10^8$ K adopted on the right panel.
{\it Right panel}: density dependence of the neutrino emissivity
through the NS core at the given $T$ for nonsuperfluid matter
(thick solid line), for  
proton superfluidity 1p or 2p (thin solid or dashed lines),
or with added neutron superfluidity 1nt or 3nt (dots).  
}
\label{fig1}
\end{figure}
%%%%%%%%%%%%%%%%%%%%%%%%%%%%%%%%%%%%%%%%%%%%%%%%%%%%%%%%%%%%%%%

The code includes the effects of nucleon superfluidities
of three types: singlet-state ($^1$S$_0$) pairing
of free neutrons in the inner crust 
(with the critical temperature
$T_{\rm c}=T_{\rm cns}(\rho)$); $^1$S$_0$ proton pairing
in the core ($T_{\rm c}=T_{\rm cp}(\rho)$); 
and triplet-state ($^3$P$_2$) neutron pairing 
in the core ($T_{\rm c}=T_{\rm cnt}(\rho)$).
The $T_{\rm c}(\rho)$ dependence
has been parameterized by simple equations (e.g., Yakovlev
et al., 2002b, c);
the adopted models of $T_{\rm c}(\rho)$ do not contradict to
numerous microscopic calculations (e.g., Lombardo and Schulze, 2001).

%%%%%%%%%%%%%%%%%%%%%%%%%%%%%%%%%%%%%%%%%%%%%%%
\subsection*{Nonsuperfluid Stellar Models}
%%%%%%%%%%%%%%%%%%%%%%%%%%%%%%%%%%%%%%%%%%%%%%%

For nonsuperfluid NSs, 
we have {\it two} well-known
cooling regimes, {\it slow} and {\it fast} cooling.
The slow cooling takes place in low-mass
NSs ($M< M_{\rm D}$) via neutrino emission produced mainly by
Murca process. The cooling
curves appear to be almost the same 
for all $M$ from about $1.1 \, {\rm M}_\odot$
to $M_{\rm D}$ (e.g., Page and Applegate, 1992)
being insensitive to EOS
(like the dot-and-dashed line in Figure 2 which can be
called {\it the standard basic cooling curve}).
These models of Murca-emitting nonsuperfluid NSs
cannot explain the observations of some sources,
first of all, PSR J0205+6449 and
Vela (too cold), as well as RX J0822--43 and PSR 1055--52 (too hot).
The data seem to require both, slower and faster cooling.

The fast cooling occurs 
via a powerful Durca process at $M>M_{\rm D}$.
The cooling curves are again not too sensitive to the mass and EOS.
These NSs are much colder
than the slowly cooling ones.
The transition from the slow to fast cooling takes
place in a very narrow range of $M$
because of the huge difference in the emissivities
of Murca and Durca processes, and a sharp threshold
of Durca process  
(right panel of Figure 1).
On the $T_{\rm s}^\infty-t$ diagram
some sources (first of all, Vela, Geminga, RX J1856--3754)
fall exactly in this transition zone, and therefore
could be explained if they had almost the same mass.
This unlikely assumption can be avoided by
including the effects of nucleon superfluidity.

%%%%%%%%%%%%%%%%%%%%%%%%%%%%%%%%%%%%%%%%%%%
\subsection*{Proton Superfluidity and
Three Types of Cooling Neutron Stars}
%%%%%%%%%%%%%%%%%%%%%%%%%%%%%%%%%%%%%%%%%%%

The observations can be explained by cooling of superfluid NSs
assuming that 
the {\it proton superfluidity
is rather strong} at $\rho \la \rho_{\rm D}$, while 
the $^3$P$_2$ {\it neutron superfluidity is rather weak}.
We start with the effects of proton superfluidity
neglecting neutron pairing.
We take two typical models of proton superfluidity,
1p and 2p, displayed on the left panel of Figure 1.
The appropriate neutrino emissivity over the NS core at
$T=3 \times 10^8$ K is
shown on the right panel. The effects of
proton superfluidity are seen to be twofold.
First, the superfluidity reduces the neutrino
emission in the outer NS core by strongly suppressing
Murca or even Durca process at not too high $\rho$.
Second, the proton superfluidity gradually dies out
with increasing $\rho$ and
removes the reduction
of fast neutrino emission.
This {\it broadens} the
threshold of fast neutrino emission, creating
{\it a finite transition zone}
(which will be denoted as $\rho_{\rm s} \la \rho \la \rho_{\rm f}$)
between slowly ($\rho \la \rho_{\rm s}$) and rapidly
($\rho \ga \rho_{\rm f})$ neutrino emitting layers.
Superfluid 2p extends deeper to the NS core and shifts the
transition zone to higher $\rho$.
The cooling curves of NSs of different masses
with proton superfluidities 1p and 2p are plotted
on the left and middle panels of Figure 2, respectively.
We see that the proton superfluidity
leads to the {\it three} representative types of cooling NSs.

%%%%%%%%%%%%%%%%%%%%%%%%%%%%%%%%%%%%%%%%%%%%%%%%%%%%%%%%%%%%%%
\begin{figure}
\centering
\epsfxsize=17.5cm
\epsffile[50 190 550 465]{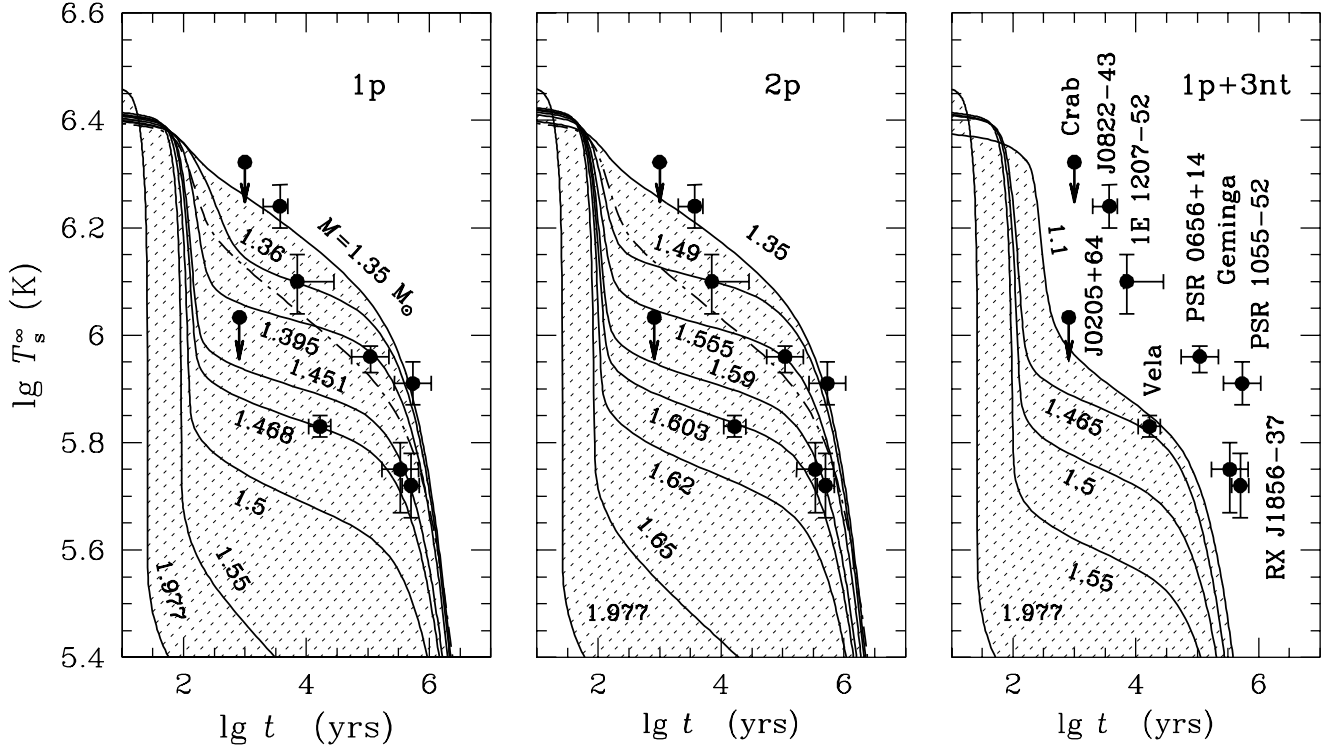}
\caption{
Observational data on nine
NSs compared with the cooling curves of NSs
of several masses $M$ for three superfluid models in the NS cores
({\it left panel}: 1p; {\it middle panel}: 2p;
{\it right panel}: 1p+3nt). Dot-and-dashed line:
cooling of a nonsuperfluid 1.35 M$_\odot$ NS. Shaded regions
are allowed by the assumed models.  
}
\label{fig2}
\end{figure}
%%%%%%%%%%%%%%%%%%%%%%%%%%%%%%%%%%%%%%%%%%%%%%%%%%%%%%%%%%%%%%%

{\it Low-mass} NSs, where $\rho_{\rm c} < \rho_{\rm s}$,
have weaker neutrino emission
than low-mass nonsuperfluid NSs; they form
a class of {\it very slowly cooling NSs}.
Their cooling curves are almost independent of
NS mass, proton superfluid model, 
and EOS in the NS cores.
These cooling curves go higher than the basic standard
cooling curve, explaining now the observations
of RX J0822--43 and PSR 1055--52.
Thus we can treat these two NSs as low-mass NSs.

{\it High-mass, rapidly cooling} NSs, where
$\rho_{\rm c} \ga \rho_{\rm f}$,
cool mainly via fast neutrino emission
from the inner core. The cooling curves are again
almost independent of $M$, EOS and proton superfluidity,
and are actually the same as for high-mass nonsuperfluid
NSs. All the observed NSs are much warmer than these ones.

{\it Medium-mass} NSs ($\rho_{\rm s} \la \rho_{\rm c} \la \rho_{\rm f}$)
show cooling which is {\it intermediate}
between very slow and fast; it
depends on $M$, EOS and proton
superfluidity. Roughly, the masses of these NSs range
from $M_{\rm D}$ to 1.55 M$_\odot$ for 1p superfluid
and from 1.4 to 1.65 M$_\odot$ for 2p superfluid.
By varying $\rho_{\rm c}$ from
$\rho_{\rm s}$ to $\rho_{\rm f}$ we obtain a family
of cooling curves which fill the space between
the curves of low-mass and high-mass NSs. We
can select those curves which explain the observations
and attribute thus certain
masses to the sources (``weigh'' NSs, Kaminker et al., 2001). 
This weighing depends on EOS and proton superfluid model
(left and mid panels of Figure 2).
We treat 1E 1207--52, Vela, PSR 0656+14, Geminga,
and RX J1856--3754 as medium-mass NSs.

%%%%%%%%%%%%%%%%%%%%%%%%%%%%%%%%%%%%%%%%%%%%%%%%%%%%%%%%%%%%%%%%%%%%%%%%%%
\subsection*{Mild Neutron Pairing in the Core Contradicts to Observations}
%%%%%%%%%%%%%%%%%%%%%%%%%%%%%%%%%%%%%%%%%%%%%%%%%%%%%%%%%%%%%%%%%%%%%%%%%%

Now we switch on $^3$P$_2$ neutron pairing in the NS core.
Microscopic theories predict this pairing to be 
weaker than the proton one, with 
$T_{\rm cnt}(\rho) \la 10^9$ K. Two models
(1nt and 3nt) are presented on the left panel of Figure 1, with
maximum $T_{\rm cnt}^{\rm max} \approx 3.4\times 10^8$ and
$8 \times 10^8$ K, respectively.
The onset of such superfluidity induces a strong
neutrino emission due to Cooper pairing of neutrons
as shown by dotted
lines on the right panel of Figure 1. The emissivity in the outer NS core
becomes higher than in a nonsuperfluid star
and accelerates the cooling. The cooling
curves for NSs with 1p proton
and 3nt neutron superfluids are shown
on the right panel of Figure 2. These NS models cool too fast
and contradict to observations of many sources.
We have checked that any {\it mild} neutron superfluidity
in the core with realistic $T_{\rm cnt}(\rho)$ profiles
and $T_{\rm cnt}^{\rm max}\sim (2 \times 10^8-2\times 10^9)$ K
contradicts to observations of at least some
hotter and older objects (independently of the
proton pairing) and should be rejected on these
grounds. A neutron superfluidity with smaller
$T_{\rm cnt}^{\rm max}$ appears at late stages
of NS evolution and has no effect on cooling of middle-aged NSs.

%%%%%%%%%%%%%%%%%%%%%%%%%%%%%%%%%%%%%%%%%%%%%%%%%%%%%%%%%%%%%
\subsection*{Very Slowly Cooling Neutron Stars and the Physics of the Crust}
%%%%%%%%%%%%%%%%%%%%%%%%%%%%%%%%%%%%%%%%%%%%%%%%%%%%%%%%%%%%%

As discussed above, the cooling of low-mass
NSs is rather robust
against uncertainties of the physics in the NS cores.
Their neutrino luminosity is exceptionally low.
Accordingly, their cooling is
{\it especially sensitive to the physics
of the NS crust}.
The cooling of these NSs (contrary to other ones)
is strongly regulated by the effects of
singlet-state neutron superfluidity
in the inner NS crusts, surface magnetic fields
and accreted envelopes as described
by Yakovlev et al.\ (2001b, 2002b) and Kaminker et al.\ (2002).
By tuning such factors one can
refine the interpretation of the observations
of RX J0822--43 and PSR 1055--52.

%%%%%%%%%%%%%%%%%%%%%%%%%%%%%%%%%%%%%%%%%%%%%%%%%%%%%
\section*{COOLING OF NEUTRON STARS WITH EXOTIC CORES}
%%%%%%%%%%%%%%%%%%%%%%%%%%%%%%%%%%%%%%%%%%%%%%%%%%%%%

At the next step we explore a hypothesis of
exotic NS cores adopting the model of
neutrino emission given by Eq.\ (\ref{Qnu}).
Quite generally, we assume the presence of
the outer NS core with slow neutrino emission,
the inner core with fast neutrino emission,
and the intermediate zone ($\rho_{\rm s} \la \rho \la \rho_{\rm f}$).
Then we
obtain three types of cooling NSs similar to those
discussed in the previous section:
low-mass NSs ($\rho_{\rm c} \la \rho_{\rm s}$)
which show slow cooling;
high-mass NSs ($\rho_{\rm c} \ga \rho_{\rm f}$)
which show fast cooling via enhanced neutrino emission
from the inner cores; medium-mass NSs
($\rho_{\rm s} \la \rho_{\rm c} \la \rho_{\rm f}$)
with intermediate cooling.

Left panel of Figure 3 displays the cooling curves
of low-mass and high-mass NSs for several
EOSs of NS cores. The EOSs
affect the neutrino emission
in the inner and outer cores. The real EOS, 
which {\it should be the same for all NSs},
is currently unknown.

The outer NS cores are thought to consist of
nucleon matter. We present two types of cooling
curves of low-mass (1.3 M$_\odot$) NSs from the previous
section. The {\it Murca} curve is the standard
basic cooling curve. The {\it brems} curve refers to 
the very slow cooling inspired by strong
proton superfluidity.
These are the {\it upper} cooling curves for NSs with
nonsuperfluid and superfluid outer cores. The cooling curves
of medium-mass NSs are lower (as shown by shading).
 
We present also three cooling curves for high-mass 
($\approx 2$ M$_\odot$)
NSs. They are calculated (Yakovlev and Haensel, 2002)  using a simplified
toy model of cooling NSs 
with three constant rates
of fast neutrino emission, Eq.\ (\ref{Qnu}), in the inner cores:
$Q_{\rm f}=10^{27}$, $10^{25}$, and $10^{23}$
erg s$^{-1}$ cm$^{-3}$. These rates are appropriate
(Table 2) to three EOSs in the inner cores:
nucleon matter with open Durca process (curve {\it nucleon Durca};
will be about the same in the presence of hyperons),
pion-condensed matter ($\pi$-{\it cond}), and kaon-condensed (or quark)
matter ({\it K-cond}). They are three possible
{\it lowest} cooling curves. The cooling curves of medium-mass
NSs are higher (as shown by shading).
For any EOS in NS cores,
we have a highest cooling curve of low-mass NSs,
a lowest cooling curve of high-mass NSs, and
a sequence of cooling curves of medium-mass
NSs between the highest and
lowest ones. All in all, we present
the allowable values of $T_{\rm s}$ for
six different EOSs.

%%%%%%%%%%%%%%%%%%%%%%%%%%%%%%%%%%%%%%%%%%%%%%%%%%%%%%%%%%%%%%
\begin{figure}
\centering
\epsfxsize=16.5cm
\epsffile[0 435 563 660]{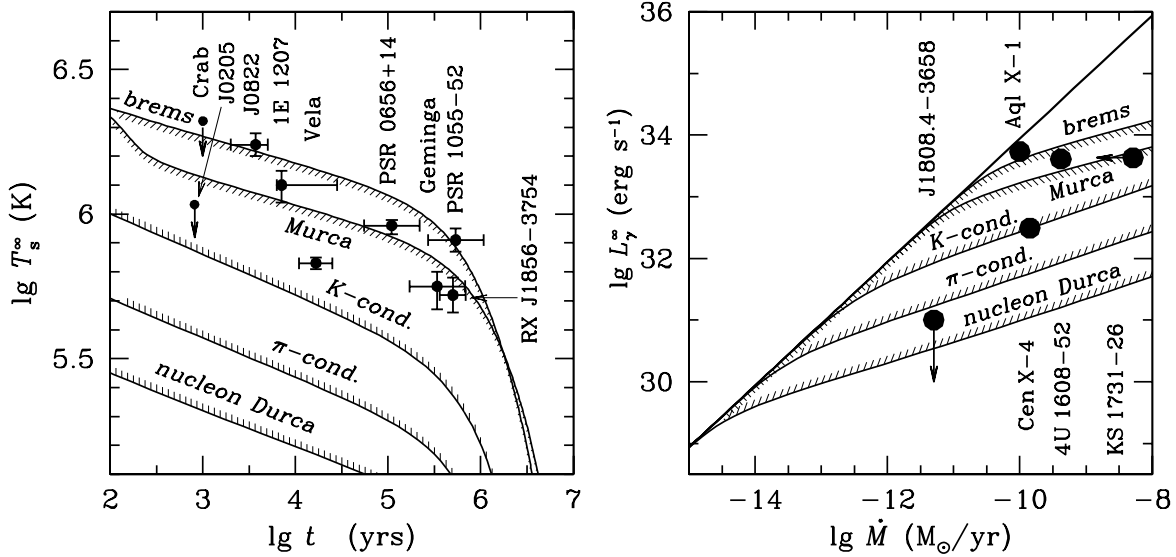}
\caption{
Observational data
confronted with theoretical curves.
{\it Left panel}:
slow cooling of low-mass NSs 
and rapid cooling of 
high-mass NSs (with nucleon
or exotic inner cores).
{\it Right panel}: heating curves of 
low-mass and high-mass accreting NSs compared
with observations of several SXRTs in quiescence.
See text for details.  
}
\label{fig3}
\end{figure}
%%%%%%%%%%%%%%%%%%%%%%%%%%%%%%%%%%%%%%%%%%%%%%%%%%%%%%%%%%%%%%%

As discussed in the previous section, curve {\it brems}
is in better agreement with the observations 
than {\it Murca}, for low-mass NSs.
All three EOSs in the inner NS cores
({\it nucleon Durca}, $\pi$-{\it cond}, {\it K-cond})
do not contradict to observations (as concluded by a number
of authors, e.g., by Page, 1998a, b). Evidently,
it would be most interesting to discover colder
NSs to constrain these EOSs.

A rather uniform scatter of the observational
points suggests the existence of the representative
class of medium-mass NSs. Their mass range is sensitive
to the position and width of the transition layer
between the slow and fast neutrino emission zones.
Unfortunately, these parameters cannot be
strongly constrained (Yakovlev and Haensel, 2002) 
from the current data.

%%%%%%%%%%%%%%%%%%%%%%%%%%%%%%%%%%%%%%%%%%%%%%%%%%%%%%%%%%%%%%
\section{THERMAL STATES OF TRANSIENTLY ACCRETING NEUTRON STARS}
%%%%%%%%%%%%%%%%%%%%%%%%%%%%%%%%%%%%%%%%%%%%%%%%%%%%%%%%%%%%%%

Now we discuss the thermal states of accreting
NSs in soft X-ray transients
(SXRTs). SXRTs undergo (e.g., Chen et al., 1997)
the periods of outburst activity
(days--months) superimposed with
quiescent periods (months--decades).
Their activity is most probably regulated
by accretion from disks around the NSs.
In quiescence,
the accretion is switched off or suppressed. Nevertheless,
some sources show rather intense
thermal radiation indicating that NSs are sufficiently
warm. It is quite possible (Brown et al., 1998)
that these NSs are warmed up by deep crustal heating
(Haensel and Zdunik, 1990) produced by nuclear transformations
in the accreted matter while this matter sinks into
the inner NS crust (to densities $\rho \ga 10^{11}$ g cm$^{-3}$)
under the weight of newly accreted material.
The total energy release is about 1.45 MeV per
accreting baryon, and the total heating power
is $L_{\rm dh} \approx 8.74 \times 10^{33}  \, 
\dot{M}/(10^{-10} \, {\rm M}_\odot \; {\rm yr}^{-1})$
erg s$^{-1}$, where $\dot{M}$ is the mass accretion rate.
The heat is spread over the NS by the thermal conductivity
and radiated away by surface photon emission and
neutrino emission from the NS interior. Generally,
the surface temperature depends on the NS internal
structure.

NSs in SXRTs are thermally inertial objects with the thermal
relaxation times $\sim 10^4$ yr (Colpi et al., 2001).
Their thermal states do not `feel' transient variations
of the accretion rate. Thus a thermal state can be
determined in the steady-state approximation by solving
the thermal balance equation: $L_{\rm dh}(\dot{M})=L_\nu^\infty
(\widetilde{T})+L_\gamma^\infty$ (cf.\ Eq.\ (\ref{therm-balance})),
where $\dot{M} \equiv \langle \dot{M} \rangle$
means the time-averaged accretion rate.
A solution gives {\it a heating curve}, $L_\gamma^\infty(\dot{M})$,
or, equivalently, $T_{\rm s}^\infty(\dot{M})$.
The heating curves of accreting NSs are closely
related to the cooling curves of isolated 
NSs (e.g., Colpi et al., 2001; Yakovlev et al., 2002a)

On the right panel of Figure 3 we present
the limiting heating curves
calculated (Yakovlev et al., 2002a)
using the toy model of NS thermal structure (Yakovlev
and Haensel, 2002). 
They are analogous to the limiting
cooling curves on the left panel.
Two upper curves ({\it brems} and {\it Murca})
are the heating curves of low-mass 
1.16 M$_\odot$ NSs ($Q_{\rm s}=3 \times 10^{19}$
and $10^{21}$ erg cm$^{-3}$ s$^{-1}$ in Table 1, respectively).
Three lower curves {\it nucleon Durca}, $\pi$-{\it cond} and {\it K-cond} 
are the heating curves of 2 M$_\odot$ NSs.
For any EOS of NS interiors, we
have its own upper heating curve and the lower one,
and a family of heating curves of medium-mass NSs
which fill the space between the upper and lower curves.

These results are confronted with observations of five SXRTs.
The data are the same as taken by Yakovlev et al.\ (2002a).
We treat $L_\gamma^\infty$ as the thermal surface
luminosity of SXRTs in quiescence, and take
the values of $L_\gamma^\infty$ for Aql X--1, Cen X--4,
4U 1608--552, KS 1731--26, and SAX 1808.4--3654
from Rutledge et al.\ (2002, 2000, 1999), Wijnands et al.\ (2002),
and Campana et al.\ (2002), respectively. 
The value of $\dot{M}$ for KS 1731--26
is most probably an upper limit. No quiescent   thermal
emission has been detected from SAX J1808.4--3658, 
and we present the established upper limit of $L_\gamma^\infty$.
Since the data are rather uncertain we plot the
observational points as thick dots. 

As seen from Figure 3,
we can treat NSs in 4U 1608--52 and Aql X--1 as low-mass
NSs with superfluid cores.
NSs in Cen X--4 and SAX J1808.4--3658 seem to require the enhanced 
neutrino emission and are thus more massive.
The status of NS in KS 1731--26 is less certain because
of poorly determined $\dot{M}$; it may also require
enhanced neutrino emission.
Similar conclusions have been made
by a number of authors (cited in Yakovlev et al., 2002a) 
with respect to some of these sources.

Let us disregard the SAX source for a moment.
The observational point for Cen X--4
lies above (or~near) all three limiting heating curves
for massive NSs. Thus we can treat
the NS in Cen X--4 either
as high-mass NS (with kaon-condensed or quark core) or as
medium-mass NS (with pion-condensed,
or nucleon Durca core).
If further observations confirm the current status of
SAX J1808.4--3658,
then we will have the only choice
to treat this NS as a high-mass NS with the
nucleon core (and the NS in Cen X--4 as medium-mass NS
with the nucleon core). This would disfavor the hypothesis
on exotic phases of matter in NS cores. 

%%%%%%%%%%%%%%%%%%%%%%%%%%%%%%%%%%%%%%%%%%%%%%%%%%
\section{CONCLUSIONS}
%%%%%%%%%%%%%%%%%%%%%%%%%%%%%%%%%%%%%%%%%%%%%%%%%%

The theory of NS cooling is very flexible and
provides many successful cooling scenarios.
If thermal states of NSs in SXRTs
are determined by deep crustal heating, the theory
and observations of isolated NSs and SXRTs can be analyzed
together.  

Disregarding the observations of SAX J1808.4--3658,
the data on isolated NSs and SXRTs imply that 
({\it a}) an enhanced neutrino emission operates in massive NSs
but its nature is unknown (nucleon Durca, pion- or
kaon-condensates, quarks?);
({\it b}) representative class of 
medium-mass NSs is available;
({\it c}) the position and width of the transition layer
between slowly and rapidly cooling layers
in the NS cores are uncertain.
If the observations of SAX J1808.4--3658 are relevant
to the present analysis, they are in favor of
nucleon models of dense matter with open Durca, 
and they disfavor the models of exotic matter.  

Some physical models of NS interiors
contradict to observations, e.g., the existence
of mild $^3$P$_2$ neutron superfluidity with
the maximum $T_{\rm cn}(\rho)$ in the range from
$2 \times 10^8$ to $2 \times 10^9$ K.

A search for new very cold and very hot
NSs would be useful. Very cold 
NSs would disfavor the models of exotic matter. 

\section*{ACKNOWLEDGMENTS}
We are indebted to George Pavlov for providing the data on PSR 1055--52.
The work was partly supported by RFBR
(grants No.\ 02-02-17668 and 03-07-90200).

\bibliographystyle{natbib} %% Added 21-11-02
\bibliography{bibfile}     %% Added 21-11-02

E-mail: yak@astro.ioffe.rssi.ru

\end{document}